%% file: paper.tex
\documentclass[10pt,sigconf,letterpaper,nonacm]{acmart}

\usepackage{graphicx}
\usepackage[labelformat=simple]{subcaption}
\usepackage{changepage}
\usepackage{multirow}
\usepackage{paralist}
\usepackage{rotating}
\usepackage{pifont}
\usepackage{tcolorbox}
\usepackage{array}
\usepackage{fontawesome5}
\usepackage{array, makecell}
\usepackage{hhline} 
\usepackage{svg}
\usepackage{float}

\newcommand{\parheading}[1]{\noindent{}\textbf{{#1}}}
\newcommand{\eg}{\textit{e.g.,}}
\newcommand{\ie}{\textit{i.e.,}}
\newcommand{\etc}{etc.}
\newcommand{\etal}{et al.}
\newcommand{\wrt}{w.r.t.}

\begin{document}

\newcommand{\diffaudit}{\textsc{DiffAudit}}

\title{{\diffaudit}: Auditing Privacy Practices of Online Services for Children and Adolescents}

\author{Olivia Figueira}
\affiliation{
  \institution{University of California, Irvine}
  \city{}
  \state{}
  \country{}
}
\author{Rahmadi Trimananda}
\affiliation{
  \institution{University of California, Irvine}
  \city{}
  \state{}
  \country{}
}
\author{Athina Markopoulou}
\affiliation{
  \institution{University of California, Irvine}
  \city{}
  \state{}
  \country{}
}
\author{Scott Jordan}
\affiliation{
  \institution{University of California, Irvine}
  \city{}
  \state{}
  \country{}
}

\input{0_abstract}

\maketitle

\input{1_intro}
\input{2_background_relatedwork}

\input{3_methodology}
\input{4_results}

\input{5_discussion}
\input{6_conclusion}

\input{7_acknowledgements}

\bibliographystyle{ACM-Reference-Format}
\bibliography{refs}

\input{appendix}

\end{document}

%% file: 0_abstract.tex
\begin{abstract}
Children's and adolescents' online data privacy are regulated by laws such as the Children's Online Privacy Protection Act (COPPA) and the California Consumer Privacy Act (CCPA). Online services that are directed towards general audiences (\ie{} including children, adolescents, and adults) must comply with these laws. 
In this paper, first, we present {\diffaudit}, a platform-agnostic privacy auditing methodology for general audience services.
{\diffaudit} performs differential analysis of network traffic data flows to compare data processing practices (i) between child, adolescent, and adult users and (ii) before and after consent is given and user age is disclosed.
We also present a data type classification method that utilizes GPT-4 and our data type ontology based on COPPA and CCPA, allowing us to identify considerably more data types than prior work.
Second, we apply {\diffaudit} to a set of popular general audience mobile and web services and observe a rich set of behaviors extracted from over 440K outgoing requests, containing 3,968 unique data types we extracted and classified.
We reveal problematic data processing practices prior to consent and age disclosure, lack of differentiation between age-specific data flows, inconsistent privacy policy disclosures, and sharing of linkable data with third parties, including advertising and tracking services.
\end{abstract}

%% file: 1_intro.tex
\section{Introduction}\label{intro} 
The Children's Online Privacy Protection Act (COPPA) in the United States (US) regulates online service providers' collection and sharing of personal information about children under the age of 13~\cite{us_federal_trade_commission_childrens_2023}. With the increased presence of both children and adolescents online~\cite{brooke_auxier_parenting_2020, emily_a_vogels_teens_2022}, more recent privacy legislation such as the California Consumer Privacy Act (CCPA)~\cite{ccpa_office_of_the_attorney_general_2023} have incorporated additional privacy protections, akin to COPPA, for users under the age of 16. For example, CCPA forbids businesses from selling and sharing the personal information of users younger than 16, unless they have received consent.
Additionally, online services, such as social media and gaming platforms, have been under increased scrutiny for their lack of safety and privacy for young users, prompting proposals for more stringent laws and regulations in the US~\cite{nyt_kosa_congress_2024, nyt_child_safety_hearing_2024, kalhan_rosenblatt_florida_2024, ap_georgia_ban_2024}. Methodologies for auditing data processing done by online services for young users can inform and enforce such regulations. 

In this paper, we develop an auditing methodology to analyze the data collection and sharing practices of online services for young users and reveal potential violations of privacy laws.
More specifically, we study online services directed towards general audiences, which include children, adolescents, and adults, and that allow users to make an account and disclose their age.
We identified the following services out of the top 100 most popular apps on the Google Play Store: Roblox~\cite{roblox_2023_site}, Minecraft~\cite{minecraft_2023_site}, TikTok~\cite{tiktok_2023_site}, YouTube~\cite{youtube_2023_site} and YouTube Kids~\cite{youtubekids_2023_site}, Duolingo~\cite{duolingo_2023_site}, and Quizlet~\cite{quizlet_2023_site}.
When creating an account on such online services, users are prompted to enter their age, which should impact the data collection and sharing that is done based on whether the user is a child (younger than 13), adolescent (between 13 and 16), or adult (older than 16), which are grouped in this way based on COPPA and CCPA.
Also, upon creating an account and agreeing to the privacy policy and terms, consent to the services' disclosed data processing practices is given, either by the parent or by the user themselves depending on the user's age. 
Given the age-specific regulations in COPPA and CCPA, we expect well-behaved online services for general audiences to perform different data collection and sharing practices when users are logged in vs. out, as well as across different age groups.

While prior work has developed frameworks and tools for privacy auditing and measurement for several platforms, such as mobile, browsers, and virtual reality~\cite{irwin_reyes_wont_2018, rahmadi_trimananda_ovrseen_2022, steven_englehardt_online_2016, anastasia_shuba_antmonitor_2017}, to the best of our knowledge, none have audited (i) general audience services (ii) in the context of both COPPA and CCPA for children and adolescents and (iii) their behaviors before and after providing consent and user age. 
Towards this goal, we make two contributions: (1) we design and implement the {\diffaudit} auditing methodology, and (2) we apply it to a set of popular general audience services and report our findings.

First, {\bf we design  {\diffaudit} -- a methodology} for auditing the practices of general audience services and comparing data collection and sharing practices across different user age groups, \ie~ children, adolescents, and adults. {\diffaudit} has the following characteristics:

\begin{compactitem}[$\bullet$]
    \item {\em Platform-Agnostic Differential Analysis}: {\diffaudit} solely relies on the availability of network traffic collected when using a service on the end device (\eg{} a mobile app, a browser, \etc{}) for different age group profiles. We perform differential analysis to compare data flows (defined as {\em <data type category, destination>}) 
    observed across age groups, as well as before and after consent is given and user age is disclosed. Destinations are also categorized as first/third parties and advertising and tracking services (ATS) using block lists.

    \item {\em Data Type Classification}: We develop a novel data type classification method using OpenAI's GPT-4 \cite{gpt-4_2023} to classify data types extracted directly from network traffic into categories. Our contribution includes our {\em data type ontology} rooted at the COPPA and CCPA definitions of identifiers and personal information~\cite{coppa_defn_childrens_2013, ccpa_defn_california_2018} (see Table~\ref{table:datatype_categories} in Appendix). Our method outperforms previously used methods both in accuracy and coverage, which we discuss in Section~\ref{sec:methods_data_types}. Prior work~\cite{irwin_reyes_wont_2018, rahmadi_trimananda_ovrseen_2022, anastasia_shuba_antmonitor_2017, yihang_song_privacyguard_2015, jingjing_ren_recon_2016, haojian_jin_why_2018, abbas_razaghpanah_apps_2018, benjamin_andow_actions_2020, trung_tin_nguyen_share_2021} focused on data types largely limited to identifiers, whereas our ontology also includes behavioral data types and is rooted in privacy laws rather than only extracted from the network traces.
\end{compactitem}

Second, \textbf{we apply {\diffaudit}} to popular general audience services on both their mobile and website platforms. In the network traffic dataset we collected, we observed more than 440K outgoing requests, with 964 unique domains and 326 unique eSLDs contacted, and we extracted 3,968 unique data types from the packets and 5,508 unique data flows. In all the services we audited with {\diffaudit}, we found practices for children and adolescent users that are concerning for COPPA and CCPA compliance, including the following: 

\begin{compactitem}[$\bullet$]

    \item {\em Data Processing Before Consent:} All services collected users' identifiers and personal information while logged out, and all but one of the services shared users' identifiers and personal information with third party ATS while logged out. This raises concerns regarding compliance due to a lack of consent and knowledge of user age.

    \item {\em Differential Auditing:} All but one of the services engaged in data processing practices that were not disclosed in their privacy policy, including while users are logged in and out, which is problematic for transparency and compliance. All general audience services we studied engaged in similar data processing practices for both young and adult users.

    \item {\em Platform Differences:} We observed some data flows that are unique to the mobile app and website platforms. We found that the observed data flows unique to the mobile apps were all related to sharing data with third parties, including personal and device identifiers, geolocation, and app usage behaviors. This reveals some potential internal differences in services' mobile vs. website platforms. 

    \item {\em Data Linkability}: All but one of the services shared linkable data (\ie{} identifiers and personal information shared with the same third parties) with both ATS and non-ATS third party domains, both before and after consent. These services also sent the linkable data to similar destination domains, without much differentiation between age groups.

\end{compactitem}

The rest of the paper is structured as follows. Section~\ref{sec:background_relatedwork} provides background and discussion of related work. Section~\ref{methodology} presents the {\diffaudit} methodology, including: the network traffic collection (Section~\ref{net_traffic}), the construction of data flows, and the data type classification method (Section~\ref{post_process}). Section~\ref{results} presents the results of applying {\diffaudit} to popular online services. Section~\ref{discussion} discusses implications of our findings and recommendations for service providers and regulators. Section~\ref{conclusion} concludes the paper. 
The Appendices include the Ethics section, our proposed data type ontology (Table \ref{table:datatype_categories}), and additional details.

%% file: 2_background_relatedwork.tex
\section{Background \& Related Work}\label{sec:background_relatedwork}

Section~\ref{law_info} discusses how COPPA and CCPA motivate the design of {\diffaudit} and Section~\ref{serv_select} discusses general audience services that we investigate. Sections~\ref{sec:related_work_auditing} and ~\ref{sec:related_work_children_privacy} discuss related work on privacy auditing and children's and adolescents' privacy, respectively.

\subsection{COPPA \& CCPA}\label{law_info}
COPPA and CCPA regulations for users under 16 motivate the design of our auditing framework. COPPA applies to services that are directed towards children, which includes services directed towards general audiences composed of users of all ages, including children, adolescents, and adults. 
According to COPPA, a ``Web site or online service directed to children means a commercial Web site or online service, or portion thereof, that is targeted to children''\footnote{16 C.F.R. § 312.2}, which may be judged based on features and content of the service (\ie{} ``subject matter, visual content, use of animated characters or child-oriented activities and incentives, music or other audio content''\footnote{16 C.F.R. § 312.2(1)}) and knowledge of children using the service~\cite{ftc_complying_2020}. 
COPPA prohibits such online service providers from collecting any personal information prior to knowing any given user's age, since the audience could contain children under 13, and even once they know the user's age, they require opt-in consent from the child's parent or guardian to collect and share such personal information, except for activities for the support of internal operations~\cite{us_federal_trade_commission_childrens_2023}.

Under CCPA, businesses are not allowed to sell nor share users' personal information for users younger than 16 years old without the user's prior affirmative authorization (\ie{} opt-in consent). 
For users under 13 years old, businesses are not allowed to sell nor share users' personal information without their parent's or guardian's opt-in consent, similar to COPPA~\cite{ccpa_teenlaw_california_2018}. 
Additionally, CCPA states, ``A business that willfully disregards the consumer’s age shall be deemed to have had actual knowledge of the consumer’s age''\footnote{CAL. CIV. Code § 1798.120(c)}.
While some of these terms (\ie{} willfully disregard, actual knowledge) are not explicitly defined, this statement can be interpreted to mean that online services should not be selling nor sharing users' data at all until they determine their age and obtain opt-in consent. 
Note that we do not audit the adult data flows against CCPA in this work, rather we only audit the child and adolescent data flows and use the adult data flows for comparison.

General audience services should wait until they determine the user's age and obtain consent, either from the user or their parent depending on their age, to perform the appropriate data collection/sharing under COPPA and CCPA. 
Thus, we develop an auditing methodology, {\diffaudit}, for general audience services based on network traffic to both investigate whether user data is being collected/shared prior to consent and age disclosure (\ie{} while logged out) and how similar the data collection/sharing practices are for different user age groups (\eg{} data collected/shared for child vs. adult and adolescent vs. adult accounts). Moving forward, we will use the term ``logged out'' to refer to the state prior to user consent and age disclosure.

One of our contributions is that we defined a data type ontology using the definitions of identifiers and personal information from COPPA and CCPA~\cite{coppa_defn_childrens_2013, ccpa_defn_california_2018}. See a brief description in Section~\ref{sec:methods_data_types} and details in Table~\ref{table:datatype_categories} in Appendix, due to lack of space.

\subsection{General Audience Services}\label{serv_select}
With our auditing goals in mind, we establish the following criteria for the types of services we wish to audit with this framework: (i) the services should target general audiences, \ie{} including children, adolescents, and adults, and (ii) the services should prompt users to make an account so that we can disclose the user's age and provide consent to the behaviors disclosed in the privacy policy.
To select services to audit, we searched through the top-100 most popular games and apps on the Google Play Store and manually inspected each app's privacy policy to determine the target audience and whether the app fit our criteria. At the time of this study, the following were the only services that met these criteria: 
Duolingo \cite{duolingo_2023_site},
Minecraft \cite{minecraft_2023_site} (owned by Microsoft), 
Quizlet \cite{quizlet_2023_site}
Roblox \cite{roblox_2023_site}, 
TikTok \cite{tiktok_2023_site}, and
YouTube \cite{youtube_2023_site}, which also includes YouTube Kids \cite{youtubekids_2023_site}. 
Roblox and Minecraft are gaming platforms, TikTok and YouTube are social media and content platforms, and Duolingo and Quizlet are educational services. We investigate each service both on the mobile and the web platforms, in addition to desktop apps for Roblox and Minecraft.

These are all important services: the websites of these six services are among the most popular on the top 1M Tranco list at the time this work was conducted (Fall 2023):  Roblox, TikTok, and YouTube were among the top 100~\cite{LePochat2019,trancowebsite}. Furthermore, the corresponding apps for these services have been cumulatively downloaded more than 12 billion times on the Google Play Store~\cite{googleplaystore} and have received close to 280 million reviews around fall 2023. 

All selected services are directed towards children, adolescents, and adults according to their terms of service, and thus both COPPA and CCPA apply. For example, YouTube has both the regular YouTube service, which we audit based on CCPA since it allows adolescents, as well as YouTube Kids, which we audit based on both CCPA and COPPA. 
The selected services allow users of all ages to access their services, with some services requiring active parental consent for users under 13, such as requiring a parent's email in the account creation process.

\subsection{Online Privacy Analysis \& Auditing}\label{sec:related_work_auditing}
Prior work has developed mechanisms for analyzing data collection and sharing practices across diverse platforms through static, dynamic, or network traffic analysis and has audited such practices against privacy policies and laws. However, few have focused specifically on children's and adolescents' privacy. There exist many tools for privacy analysis and auditing for mobile devices and apps that are based on network traffic analysis to investigate data collection and sharing practices as well as presence of ATS~\cite{haojian_jin_why_2018, abbas_razaghpanah_apps_2018, jingjing_ren_recon_2016, yihang_song_privacyguard_2015, anastasia_shuba_antmonitor_2017}. Others have used dynamic taint analysis to monitor real-time data sharing on Android~\cite{william_enck_taintdroid_2014} as well as static analysis to analyze third-party tracking on the Google Play Store~\cite{reuben_binns_third_2018}. For websites and browsers, prior work has developed privacy measurement tools for network traffic and user advertising consent preferences~\cite{steven_englehardt_online_2016, zengrui_liu_opted_2022}. 
Researchers have also developed frameworks to investigate privacy of emerging platforms, such as virtual reality device app behaviors~\cite{rahmadi_trimananda_ovrseen_2022}, personal information exposure and prevalence of ATS in smart TVs~\cite{janus_varmarken_tv_2020, hooman_mohajeri_moghaddam_watching_2019}, and smart speaker tracking behaviors~\cite{umar_iqbal_tracking_2023}. 
Additionally, prior work has studied consent in third-party tracking and data sharing in Android apps~\cite{konrad_kollnig_fait_2021, trung_tin_nguyen_share_2021}.

Specifically related to children's privacy, researchers have analyzed privacy practices of mobile apps intended for children~\cite{irwin_reyes_wont_2018}, risky content for children in voice assistants~\cite{tu_le_skillbot_2022}, and targeted advertising towards children online \cite{tinhinane_medjkoune_marketing_2023}. Others have audited the privacy and security of parental control applications~\cite{suzan_ali_betrayed_2020, alvaro_feal_angel_2020}, algorithmic personalization on TikTok for young users~\cite{martin_hilbert_bigtech_2024}, and IoT smart toys for children~\cite{gordon_chu_security_2019}. While these works revealed privacy issues and advanced transparency and privacy, more investigation is needed as technology continues to evolve and become increasingly accessible to children and adolescents, introducing new risks.

The closest to our work is that of Reyes \etal{}~\cite{irwin_reyes_wont_2018}, which analyzed data collection and sharing practices of Android mobile apps specifically for children \wrt{} COPPA using dynamic analysis. Our work differs in that (i) we specifically investigate general audience services, which have not been previously studied, (ii) on both mobile and website platforms, and (iii) we apply a differential analysis approach based on age and consent status (\ie{} while logged in and out) in context of both COPPA and CCPA. In addition, our data type classification method expands beyond prior work and allows us to identify considerably more data types.

\subsection{Children's \& Adolescents' Privacy}\label{sec:related_work_children_privacy}
The research community has also investigated how users, including children, adolescents, and parents, approach children's and adolescents' privacy from different perspectives.

Prior works have studied how young users and parents understand and make decisions about online privacy, such as privacy conceptualizations across technologies in general~\cite{jun_zhao_i_2019, kaiwen_sun_they_2021, leah_zhang-kennedy_nosy_2016, priya_kumar_no_2017}, virtual reality systems~\cite{elmira_deldari_investigation_2023}, social media~\cite{cristiana_s_silva_privacy_2017, danah_boyd_why_2011, hyunjin_kang_teens_2022, nico_ebert_creative_2023}, and smart toys~\cite{emily_mcreynolds_toys_2017, noah_apthorpe_evaluating_2019}. Researchers have also investigated advertising directed towards children and adolescents through content analysis and studies of user perceptions~\cite{cami_goray_youths_2022, marisa_meyer_advertising_2019, xiaomei_cai_advertisements_2008, xiaomei_cai_online_2013}.
Also, researchers have studied the relationship between parents and their children regarding technology usage and privacy management, including experiences with parental controls~\cite{arup_kumar_ghosh_safety_2018, peiyi_yang_towards_2023} and parental oversight strategies~\cite{mamtaj_akter_parental_2022, maria_grazia_lo_cricchio_parental_2022, pamela_wisniewski_parental_2017, phoebe_k_chua_what_2021}. Prior works have also studied age assurance and parental consent mechanisms from legislative and policy-making perspectives~\cite{adam_d_thierer_social_2007, scott_babwah_brennen_keeping_2023, simone_van_der_hof_we_2022}, through literature reviews~\cite{chelsea_jarvie_are_2021}, or via small-scale analyses of mechanisms currently in use on popular platforms~\cite{liliana_pasquale_digital_2022}. 

Prior works have also developed tools to both protect user privacy and increase transparency, such as privacy protections for mobile apps~\cite{bing_hu_protecting_2015, france_belanger_pocket_2013}, smart toys ~\cite{siew_yong_risk_2011} and facial recognition technologies~\cite{alem_fitwi_minor_2020, reneta_p_barneva_study_2017}, as well as design insights for privacy labels and warnings \cite{john_dempsey_children_2022, max_van_kleek_better_2017}. 
The research community has also studied barriers faced by developers in designing applications for children~\cite{anirudh_ekambaranathan_money_2021, ekambaranathan_how_2023} and has proposed design guidelines for children's privacy~\cite{liccardi_can_2014, ge_wang_12_2023, ge_wang_informing_2022, priya_kumar_co_designing_2018}. 

We contribute to this body of work by demonstrating that services meant for children and adolescents continue to behave in privacy-perverse ways. Additionally, many of the observed behaviors are difficult to protect against with existing tools for children's and adolescents' privacy.

%% file: 3_methodology.tex
\begin{figure*}[ht!]
    \centering
    \includegraphics[width=1\textwidth]{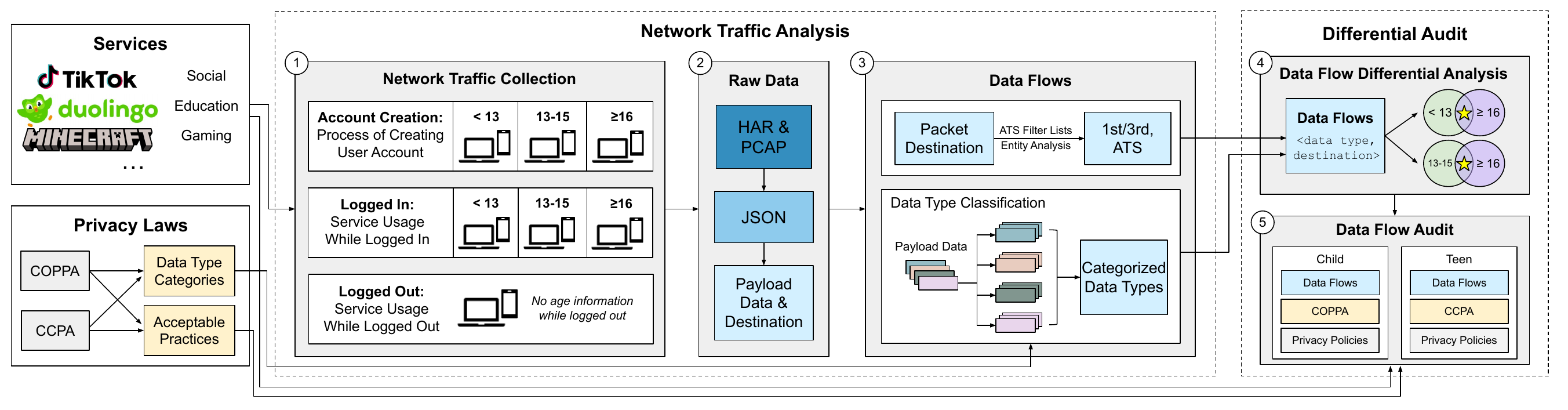}
    \begin{minipage}{\linewidth}
      \caption{{\diffaudit} Framework Overview. 
      {{\normalfont{First, we select the general audience services for auditing. Next, we perform}}
      {network traffic analysis:} {\normalfont{(1) Collect network traffic in three different ways:}} {\textit{account creation trace}} {\normalfont{(we collect traffic while creating a user account),}} {\textit{logged-in trace}} {\normalfont{(we collect traffic only while logged in to each account on the website/app and throughout using the service), and}} {\textit{logged-out trace}} {\normalfont{(we collect traffic only while logged out of any account on the website/app). (2) For each trace, we convert the raw HAR (for website) or PCAP (for mobile) data to JSON and extract the payload data and destinations of the packets. (3) We construct the data flows by processing packet destinations (\ie{} 1st or 3rd party and entity analysis) and perform data type classification using GPT-4 classifier and COPPA/CCPA data type ontology. Next, we perform the}} differential audit: {\normalfont{(4) compare the data flows by age group and (5) audit the flows in context of each age group, applicable law, and information from each service's privacy policy.}}}}
      \label{fig:auditing_framework}
    \end{minipage}
    \Description[This is a block diagram visualizing all components of the {\diffaudit} framework]{This is a block diagram visualizing all components of the {\diffaudit} framework. On the left there are two blocks, one called "Services" for the services selected to audit and one called "Privacy Laws". The "Services" block lists some of the services we have selected to audit. The "Privacy Laws" block as two blocks for COPPA and CCPA inside, both pointing to two other boxes labeled "Data Type Categories" and "Acceptable Practices", which come from the regulations. In the middle is a main block called "Network Traffic Analysis", which has three components: Network Traffic Collection, Raw Data, and Data Flows. The "Network Traffic Collection" block details the three categories of traces we do to collect traffic. The "Raw Data" block shows three blocks for post-processing the network traffic, which is to convert the HAR and PCAP files to JSON and then to extract the payload data and destination data. The "Data Flows" block has two blocks inside, one to show that we extract 1st/3rd party and ATS data from the destinations of the packets and another to show that we conduct data type classification from the traffic data. This box points to the second main block besides the "Network Traffic Analysis", which is called "Differential Audit". The "Differential Audit" block has two blocks inside, one showing that we do differential analysis on the data flows we constructed in the previous step, and another that shows that we take the privacy policy and law information from the "Privacy Laws" and "Services" blocks to audit the data flows in context.} 
\end{figure*}

\section{The {\diffaudit} Methodology}\label{methodology}

In this section, we describe how we collect and analyze the outgoing network traffic from the mobile apps and websites while using the services of interest. In Section~\ref{net_traffic}, we describe the setup and process for network traffic collection, including collecting traffic while logged out and for each user age group while logged in. In Section~\ref{post_process}, we describe how we post-process the network traffic traces to extract the data flows, including our data type classification method to categorize extracted data types and how we categorize the packet destinations.

\subsection{Network Traffic Collection}\label{net_traffic}
Our auditing framework, as shown in Figure \ref{fig:auditing_framework}, relies on network traffic analysis. To be able to compare network traffic transmitted on each service for different age categories, we created three profiles on each service, one for each of the three age categories: child (younger than 13), adolescent (at least 13 and younger than 16), and adult (16 and older). We collected network traffic for each profile on the mobile app and website for each service, amounting to three traces per app and per website. In addition, we collected three types of network traces to compare the data processing practices before, during, and after account creation: (1) {\em Account creation trace:} We collected data throughout the entire process of creating an account (which includes providing user age and consenting to the privacy policy and terms of service). (2) {\em Logged-in trace:} We collected data only during service usage, meaning we logged in to each account and then began collecting network traffic only while logged in (provides user age due to being logged in). (3) {\em Logged-out trace:} We collected data only while logged out, so there are no age-specific traces in this category (\ie{} consent and age have not been given/disclosed). We examine the default state of these services, meaning we did not opt in nor out of any data collection/sharing for any of the age categories beyond what was consented to in the privacy policies.

While there exist tools for automated interaction, we chose to use manual interaction because it enables exhaustive exploration of the services and all their features, allowing for more in-depth exploration of each service. Apps and websites often change their layouts and services depending on the age of the user currently signed in, such as having different features for child versus adult accounts, which complicates automated interaction methods. Manual interaction also resembles realistic service usage and has been used in prior work for privacy auditing tasks for this reason \cite{rahmadi_trimananda_ovrseen_2022, jingjing_ren_recon_2016, janus_varmarken_tv_2020, haojian_jin_why_2018}\footnote{For the six popular general audience services we study in depth in this paper, manual interaction is possible and preferred for the reasons mentioned above. However, even if the interaction with the services becomes automated in the future, the rest of the {\diffaudit} framework still holds.}. For each kind of trace, we attempted to utilize every feature available in the service, and the account creation and logged-in traces should be a minimum of five minutes, whereas the logged-out trace category is usually shorter as we are limited in functionality while logged out. Thus, we have about 10-15 minutes of data collection time per age category for each service on both platforms (\ie{} about one hour total per service). Table~\ref{table:network_dataset_stats} presents the summary of our resulting network traffic dataset per service, with mobile and website data merged. We observed 964 unique domains and 326 unique eSLDs among 440,513 total outgoing packets in total.

\input{tables/traffic_statistics}

\subsubsection{\textbf{Mobile Apps}} 
To perform the network traffic collection on mobile apps, we used PCAPdroid~\cite{emanuele_faranda_pcapdroid_2023} on a Pixel 6 Android mobile device. PCAPdroid is an open source app that can track, analyze, and block the connections of other apps running on the same Android device. PCAPdroid simulates a VPN to capture the network traffic without root: it does not use a remote VPN server and data is processed locally on the device. However, to decrypt the network traffic, we had to root the device and install PCAPdroid's user certificate as a trusted certificate~\cite{emanuele_faranda_pcapdroid_2023-1}. To collect network traffic, we launch the target app, begin the trace within PCAPdroid, and use the target app to generate traffic. When we are done with the trace, we stop the collection in PCAPdroid and save the generated PCAP file to local device storage. PCAPdroid also generates a TLS key log file, which we save to local device storage and use to decrypt the generated PCAP file with Wireshark~\cite{wireshark_2023}. We clear the storage and cache of the app in the device settings between traces. 
We attempted to get as much decrypted traffic as possible by also employing certificate pinning bypass techniques using Frida~\cite{frida} along with PCAPdroid. We include all collected traffic, both encrypted and decrypted, in our analysis.

\subsubsection{\textbf{Websites}} 
To perform the network traffic collection on the website services, we used the Chrome browser~\cite{google_chrome_2023} and Chrome DevTools Network Panel~\cite{kayce_basques_chrome_2019}. We ensured the browser was reset to its default settings and cleared all browsing data (\ie{} all browsing history, cookies, site data, and cached data) between each trace. For each trace, we ensured that the ``Preserve logs'' checkbox is selected before recording to ensure all traffic is collected when new sites are opened during the same session. After completing each network trace, we export the trace as a HAR (HTTP Archive) file~\cite{jan_odvarko_har_2007}, as provided by the Network Panel. 

\subsubsection{\textbf{Desktop Apps}}
Roblox and Minecraft also provide complementary desktop apps with more features, namely the gaming features of these services. Thus, we also collected network traffic in the same manner from the desktop app versions of these services using Proxyman~\cite{proxyman_2023}, which is a man-in-the-middle system that we use to capture and decrypt traffic through SSL proxying, and we export the traces to HAR files, similarly to the website process.

\subsection{Network Traffic Post-Processing}\label{post_process}

We use PCAPdroid to generate a PCAP file and TLS key log file for each mobile app trace. 
We use the Wireshark functionality editcap~\cite{wireshark_embedding_2023} to embed the TLS keys into the PCAP file and generate a new decrypted version of the PCAP file, which we then use for the subsequent network traffic post-processing steps. We convert the HAR and PCAP files to JSON format and extract all outgoing requests from each trace, as we are interested in the data leaving our device.

\subsubsection{\textbf{Data Flows in Context}}
Starting from a raw, clear-text network traffic trace, we extract each packet's destination and data types from the packet headers and payloads, respectively. We use that information to construct {\em data flows}, defined as a pair {\em <data type category, destination>}, which is consistent with the terminology in \cite{rahmadi_trimananda_ovrseen_2022, benjamin_andow_actions_2020, hao_cui_poligraph_2023}.
We determine the appropriateness of a data flow based on the user's age and logged-in/out status (\ie{} indicating consent) in context with COPPA and CCPA. This can be thought of as a special case of appropriate information flows in the contextual integrity framework~\cite{helen_nissenbaum_privacy_2009, yan_shvartzshnaider_going_2019}.

\subsubsection{\textbf{Data Type Classification Methodology}}\label{sec:methods_data_types}

We extract key-value pairs from the JSON-structured data, and the keys serve as the raw data types, which is common in prior work. To categorize these raw data types, prior works have utilized manual inspection~\cite{rahmadi_trimananda_ovrseen_2022}, string-matching techniques using pre-determined regular expressions~\cite{anastasia_shuba_antmonitor_2017, yihang_song_privacyguard_2015}, supervised learning with natural language processing techniques~\cite{jingjing_ren_recon_2016}, or a combination of these techniques~\cite{haojian_jin_why_2018}. These prior works utilize data type ontologies, developed heuristically, and often require prior knowledge of user identifiers to directly search for in the network traffic or a precise list of strings to match for with various permutations.

\begin{table}[ht!]
  \centering
  \small
  \begin{minipage}{\linewidth}
      \caption{Data Type Categories From Our Ontology.
      {{\normalfont{Our proposed data type ontology is based on COPPA/CCPA. Data types observed in our dataset are marked with `*'.}}}}\label{table:data_type_35_categories}
    \end{minipage} 
  \begin{tabular}{l l}\hline
    \multicolumn{2}{c}{\bf{Identifiers~\cite{ccpa_defn_california_2018, coppa_defn_childrens_2013}}} \\\hline
    Name*        &   Customer Numbers \\
    Linked Personal Ids. & Login Info* \\
    Contact Info* & Device Hardware Ids.*\\
    Aliases* & Device Software Ids.*\\
    Reasonably Linkable Pers- & Device Info* \\
    onal Ids.* & \\\hline
    
    \multicolumn{2}{c}{\bf{Personal Information~\cite{ccpa_defn_california_2018, coppa_defn_childrens_2013}}}   \\\hline
    Race & Precise Geolocation \\
    Age* & Coarse Geolocation* \\
    Language* &  Location Time* \\
    Religion    & Communications\\
    Gender/Sex*  & Contacts\\
    Marital Status  & Internet Activity\\
    Military/Veteran Status & Network Connection Info* \\
    Medical Conditions  & Sensor Data\\
    Genetic Info & Products \& Advertising* \\
    Disabilities    & App/Service Usage*\\
    Biometric Info  & Account Settings* \\
    Personal History   & Service Info* \\
    Inference About Users*\\\hline
  \end{tabular}   
\end{table}

\input{tables/model_validation_tables}

In contrast, we propose a {\em{data type ontology}}, which we define based on COPPA and CCPA legal definitions of identifiers and personal information~\cite{coppa_defn_childrens_2013, ccpa_defn_california_2018}. Due to lack of space, we show an excerpt of the multi-level ontology in Figure~\ref{fig:data_labeling_gpt_diagram} and the full ontology can be found in the Appendix in Table~\ref{table:datatype_categories}.
This ontology is one of our contributions, as it is critical to our data type classification methodology. Our ontology has four levels, where level 1 consists of these two main categories (identifiers and personal information). Level 2 splits into 8 categories (\ie{} personal identifiers, device identifiers, personal characteristics, personal history, geolocation, user communications, sensors, and user interests and behaviors), which split into 35 level 3 categories, as shown in Table~\ref{table:data_type_35_categories}. Table~\ref{table:data_type_35_categories} also indicates with an `*' which data types were observed in our dataset. Level 4 contains the data types belonging to each of the 35 level 3 categories. Levels 3 and 4 are used in our data type classification methodology, as we discuss later in this section.

Prior works that conduct data type categorization primarily focus on identifiers and include a limited number of data types \cite{irwin_reyes_wont_2018, rahmadi_trimananda_ovrseen_2022, anastasia_shuba_antmonitor_2017, yihang_song_privacyguard_2015, jingjing_ren_recon_2016, haojian_jin_why_2018, abbas_razaghpanah_apps_2018, benjamin_andow_actions_2020, trung_tin_nguyen_share_2021}, with \cite{jingjing_ren_recon_2016} including more than the rest, but still only covers 12 of the 35 data types we include (\ie{} device hardware/software identifiers, name, alias, gender/sex, age, contact information, contacts, marital status, precise/coarse geolocation, and login information). 
With our data type ontology, we are able to capture considerably more data types, including behavioral data.

Our {\em{data type classification method}} is built using our data type ontology and OpenAI's GPT-4 8K context model~\cite{gpt-4_2023} with their Chat Completion API~\cite{chat_completions_api_2023}, which can take in prompts and generate textual responses. Note that this work was completed prior to OpenAI's launch of GPT-4 Turbo and other updates to their APIs\footnote{OpenAI DevDay, November 2023: https://openai.com/blog/new-models-and-developer-products-announced-at-devday}. Our approach, depicted in Figure~\ref{fig:data_labeling_gpt_diagram}, is similar to a few-shot classification framework, where we treat GPT-4 as a classifier that takes in the category labels (level 3 categories from our ontology) and data types in each category (level 4 data types from our ontology), and the model must classify our raw input texts using these categories. We had to determine what value to use for GPT-4's temperature parameter, which dictates how creative the model will be in its response and ranges from 0 to 2~\cite{create_temp_chat_completions_2023}. For values greater than 1, we observed hallucinatory responses, so we only used values between 0 to 1 in increments of 0.25. We also prompted the GPT-4 model to output a confidence level along with each output to serve as a confidence threshold for our final labels. Due to the nature of network traffic data, there are input texts that will be unclear to the model, as well as to humans doing manual labeling, such as unknown acronyms or seemingly random characters that have internal meaning known only to the app developers, and we do not want to include such low confidence guesses in the dataset. See Appendix~\ref{app:labeling_methods} for details about the model prompt.

\begin{figure}[t!]
    \centering
    \includegraphics[width=0.48\textwidth]{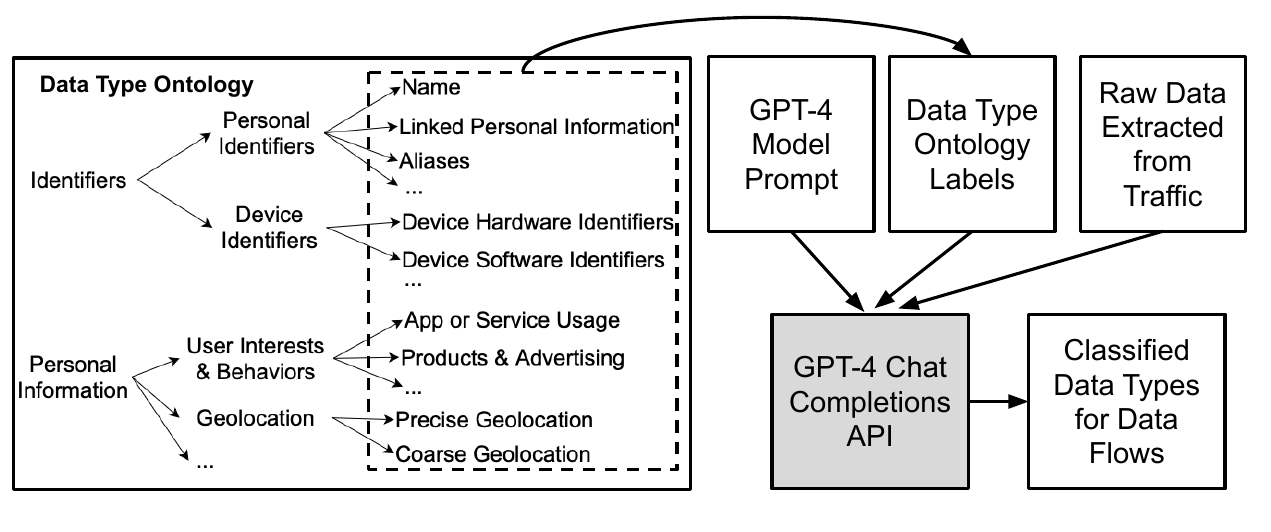}
        \begin{minipage}{\linewidth}
          \caption{Data Type Classification System Diagram. 
          {
          {\normalfont{On the left we present an excerpt of the}}
          {\em{data type ontology.}}
          {\normalfont{The data type category labels (third level) are used as labels for the classifier. We input the}}
          {\em{GPT-4 model prompt, data type ontology labels,}}
          {\normalfont{and}}
          {\em{raw data types}}
          {\normalfont{extracted from the network traffic to the}}
          {\em{GPT-4 Chat Completions API,}}
          {\normalfont{which outputs classification results.}}}}
        \label{fig:data_labeling_gpt_diagram}
    \end{minipage}    
    \Description[Block diagram visualizing all components of the data type classification framework.]{Block diagram visualizing all components of the data type classification framework. On the left is a block with an excerpt of the data type ontology, which points to a block on the right for data type ontology labels. On the right there are also two other boxes for the GPT-4 model prompt and raw data extracted from traffic. Those three blocks on the right all point to the GPT-4 Chat Completions API block, which points to another block on the right for classified data types for data flows as output.} 
\end{figure}

Considering the inherent nondeterminism of GPT-4, we build a majority-vote model where we take the majority label assigned across all the different temperature models and assign that majority label as the final label to attempt to balance model creativity, accuracy, and nondeterminism. For the majority-vote model confidence score threshold, we either compute the accuracy and number of inputs labeled based on the maximum confidence score amongst the models that assigned the majority label or we can use the average confidence score of the scores given by the models that assigned the majority label. 

We also experimented with other labeling methods, namely fuzzy string matching with TF-IDF and BERT embeddings~\cite{maarten_grootendorst_maartengrpolyfuzz_2021} and both zero- and few-shot classification with transformers~\cite{huggingface_zeroshot_2023, tunstall_efficient_2022}, but such approaches achieved poor accuracy when validated on a manually-labeled sample of our dataset (see next section). The raw network traffic includes a myriad of string formats, 
such as those that directly relate to their meaning (\eg{} ``email'', ``username''), acronyms and abbreviations of well-defined terms (\eg{} ``os'', ``rtt''), and well-defined terms concatenated with other text and/or punctuation (\eg{} ``pers\_ad\_show\_third\_part\_measurement'', ``IsOptOutEmailShown''). 
GPT-4 is better equipped to extract the context and meaning compared to previously used methods and thus achieves greater accuracy. Additionally, we do not require prior knowledge of user identifiers, allowing for a much more flexible approach.
Appendix~\ref{app:labeling_methods} provides more detail regarding the alternative classifier experiments.

We manually labeled a random sample of 10\% (n=397) of the entire dataset of extracted network traffic data types using our classification categories. We ran each of the classification approaches and computed the accuracy based on our manually-labeled sample. Besides GPT-4, the best performing alternative was fuzzy string matching with TF-IDF embeddings at 31\%, and the rest achieved even lower sample accuracy scores (zero-shot at 4\%, few-shot at 16\%, and fuzzy string matching with BERT at 18\%). We present in detail the sample accuracy results and coverage counts at different confidence levels for each GPT-4 model in Table~\ref{table:gpt4_results}. The GPT-4 models perform significantly better than the other classification approaches mentioned, and the majority-vote approaches perform better than the individual temperature GPT-4 models when taking into account the confidence thresholds. To maximize accuracy while maintaining reasonable coverage, we use the majority-average approach at the 0.8 confidence threshold in our final labeling scheme. We also manually validate a sample of the final results, as should be done with any unsupervised mechanism.

Overall, this data type classification approach offers flexibility in its design through the confidence thresholds and options regarding single temperature-based models or majority-vote models, depending on the users' preferences \wrt{} coverage, accuracy, and cost. Additionally, our method produces a set of labeled network traffic payload data that can be used to train smaller models that can be run locally instead.

\subsubsection{\textbf{Destinations}}\label{sec:methods_destinations}
To complete the data flows, we also post-process the packet destinations. To determine the fully qualified domain name (FQDN) destination for each request, we extracted the FQDN from the destination URL in each packet in the HAR and PCAP traces. Then, we extracted the effective second level domain (eSLD) using the tldextract library~\cite{john_kurkowski_john-kurkowski_2023}. Based on the eSLD, we determine the parent organization owner of this domain, using whois~\cite{richard_penman_whois_2023} and the DuckDuckGo Tracker Radar dataset~\cite{tracker_radar_duckduckgo_2023} if possible. 
The Tracker Radar dataset consists of the most commonly contacted third-party domains and has information about ownership, classification, and behavior, such as fingerprinting. 

We categorize the domains as first parties if we can match it to the name of the service (\ie{} roblox.com) or if we identify that the parent owner of the domain is the same as that of the service. Otherwise, we categorize the domain as a third party. Furthermore, we identify which domains are likely to be advertising and tracking services (ATS) using ATS block lists~\cite{big_blocklist_2023} based on the FQDN, as prior work has also done~\cite{rahmadi_trimananda_ovrseen_2022, janus_varmarken_tv_2020, benjamin_andow_actions_2020, umar_iqbal_tracking_2023}. If any of the block lists results in a block decision for a particular domain, we label that domain as an ATS and store that information in a separate label, which we use in our data flow analysis. Overall, we identify whether each domain is a first party, a first party labeled as an ATS, a third party, or a third party labeled as an ATS.

%% file: tables/traffic_statistics.tex
\begin{table}[t!]
  \centering
  \small
  \caption{Network Traffic Dataset Summary. \normalfont{The same domains and eSLDs may appear across the different services, thus their totals are based on their unique totals. YouTube includes YouTube Kids.}}
  \begin{tabular}{lcccc}\hline
    \textbf{Service}        &  \textbf{Domains} &  \textbf{eSLDs}  &   \textbf{Packets}  & \textbf{TCP Flows}   \\\hline
    Duolingo                &       122         &       69          &   60,909      &  1,466 \\\hline
    Minecraft               &       136         &       56          &   134,852     &  2,004 \\\hline
    Quizlet                 &       532         &       257         &   88,102      &  6,158 \\\hline
    Roblox                  &       152         &       24          &   103,642     &  2,302 \\\hline
    TikTok                  &       80          &       14          &   32,234      &  2,412 \\\hline
    YouTube                 &       76          &       15          &   20,774      &  226 \\\hline\hline
    \textbf{Total}          & \textbf{964}      &  \textbf{326}     &  \textbf{440,513} & \textbf{14,568}  \\\hline
    \end{tabular}
    \label{table:network_dataset_stats}
\end{table}

%% file: tables/model_validation_tables.tex
\newcolumntype{C}[1]{>{\centering\arraybackslash}p{#1}}
\newcolumntype{R}[1]{>{\raggedright\arraybackslash}p{#1}}

\begin{table*}[t]
    \centering
    \small
    \caption{GPT-4 Classification Model Sample Validation Results.
    {{\normalfont{Accuracy and coverage results are based on a random sample. The table shows total sample accuracy and per-confidence score results for each GPT-4 model varying by temperature and method.}}}}
    \begin{tabular}{C{1.8cm}C{1.2cm}C{1.2cm}C{1cm}C{1.2cm}C{1cm}C{1.2cm}C{1cm}}\hline
        \bf{Temperature}  & \multirow{2}{*}{\bf{Accuracy}} & \multicolumn{2}{c}{\bf{Confidence 0.7}} &  \multicolumn{2}{c}{\bf{Confidence 0.8}}    & \multicolumn{2}{c}{\bf{Confidence 0.9}}   \\\hhline{~~------}
        \cline{3-8}
        \bf{or Method} & & \bf{Accuracy} & \bf{Labeled} & \bf{Accuracy} & \bf{Labeled} & \bf{Accuracy} & \bf{Labeled} \\
        \hline
        \centering{0}	    &	0.72	&	0.76	&	363	&	0.82	&	292	&	0.87	&	221	\\\hline
        \centering{0.25}	&	0.74	&	0.77	&	374	&	0.84	&	294	&	0.85	&	230	\\\hline
        \centering{0.5}	    &	0.69	&	0.73	&	368	&	0.79	&	297	&	0.81	&	224	\\\hline
        \centering{0.75}	&	0.66	&	0.71	&	353	&	0.78	&	271	&	0.81	&	187	\\\hline
        \centering{1.0}	    &	0.65	&	0.70	&	356	&	0.75	&	286	&	0.78	&	201	\\\hline
        \centering{Majority-Max}	&	0.75	&	0.77	&	381	&	0.82	&	354	&	0.85	&	299	\\\hline
        \centering{Majority-Avg}	&	0.75	&	0.81	&	358	&	0.87	&	274	&	0.91	&	162	\\\hline
    \end{tabular}
    \label{table:gpt4_results}
\end{table*}

%% file: 4_results.tex
\section{Results}\label{results}

In this section, we analyze the results of our audits and discuss the findings. Section~\ref{sec:data_flow_contents} discusses data flows observed for each service, including for the logged-out traces and per-age data flows, and compares the behaviors against each service's privacy policy. Section~\ref{sec:data_linkability_results} presents the data linkability results.

\input{tables/web_vs_mobile_data_flows}

\subsection{\textbf{Differential Data Flow Auditing}}\label{sec:data_flow_contents}
Table \ref{table:platform_comparison_dataflows} shows the data flows observed across all network traces per service, with account creation and logged-in traces merged together. ``Collect'' indicates data sent to first parties, and ``share'' indicates data sent to third parties. The table indicates with different symbols whether data flows were observed on web, mobile, both, or neither, as shown in the caption. Recall that we were not able to collect all the network traffic in clear-text on the mobile apps, so the seeming lack of data flows unique to mobile is likely due to this difference.

Since there are 19 data type categories observed out of the 35 possible categories, as shown in Table~\ref{table:data_type_35_categories}, we only show the abstracted categories (level 2 from ontology) due to space, resulting in six data type categories shown in Table \ref{table:platform_comparison_dataflows}. 
For the purposes of auditing against COPPA and CCPA, the six abstracted categories suffice as they represent the categories that are covered in the laws, as shown in the full ontology in Appendix Table~\ref{table:datatype_categories}.

We analyze whether there are differences between the age-specific and the logged-out traces to determine whether the data processing is changing as we expect and whether it is appropriate for these categories. 
For the child category, if we observe any data flows among these law-protected categories either being collected or being shared (\ie{} all the collect and share columns in Table \ref{table:platform_comparison_dataflows}) that are not clearly disclosed in the privacy policy, those data flows would raise concerns and necessitate further investigation, such as by regulators. 
For the adolescent trace category, if we observe any data flows among these law-protected categories being shared with third parties, particularly ATS domains that indicate non-functional data flows, without clear disclosure in the privacy policies, those data flows would raise concerns.
Next, we will discuss the logged-out (pre-consent) category data flows, auditing results for each individual service, and data flow differences observed between the services hosted on different platforms (\ie{} web, mobile, or both).

\subsubsection{\textbf{Logged-Out Traces}}
Across Table \ref{table:platform_comparison_dataflows}, we observe many data flows in the logged-out category across all services. 
All the data types categories except for geolocation were observed being collected by first party domains while logged-out across all services (only Roblox did not collect geolocation). 
Additionally, many of the other data types were either collected by first party ATS, shared with third parties, or shared with third party ATS. 
For example, we observed device identifiers and user communications data being shared with third party ATS for all services except YouTube. 
Additionally, we observed data flows containing all data type categories being shared with third party ATS for Roblox, Duolingo, and Quizlet. 

It is concerning that personal information and identifiers are collected and/or shared prior to the user informing the service of their age and prior to giving consent through the account creation procedure, particularly for the child and adolescent users for whom services must obtain opt-in consent.
General audience services should wait until they are informed of user age and receive the appropriate consent so that they proceed with the appropriate data collection and sharing for the age category.

\parheading{\textit{Key Takeaways:}} All of the services engaged in data collection and/or sharing prior to consent and age disclosure, which is problematic for child and adolescent users for whom these services must obtain opt-in consent. All but one of the services (YouTube) was observed sharing identifiers and personal information with third party ATS while logged-out.

\subsubsection{\textbf{Age-Specific Traces \& Privacy Policy Analysis}}\label{sec:diffaudit_policy_anaysis}
Next, we will discuss the data flows observed for each individual service under audit based on Table~\ref{table:platform_comparison_dataflows} in addition to details about what data flows were disclosed in each service's privacy policy at the time of data collection (fall 2023).

\parheading{Duolingo:} We observed that all the listed data type categories, including personal and device identifiers, personal characteristics, geolocation, user communications, and user interests and behaviors, were collected by first parties and shared with third party ATS across all age categories. This is alarming both regarding the lack of difference between age groups and the data sharing particularly to third party ATS in the child and adolescent traces.

According to Duolingo's privacy policy, for users under 16 ``advertisements are set to non-personalised'' and ``third-party behavioral tracking is disabled''~\cite{duolingo_2023_privacy_policy}. However, we observed several data types being sent to third-party ATS for both the child and adolescent groups that suggest personalized ads and behavioral tracking, including both personal and device identifiers, personal characteristics, geolocation, and user interests and behaviors. We question the purpose of these data flows to third party ATS if not for behavioral tracking and personalized ads. Also, the same data type categories were shared with third party ATS for the adult trace, which further casts doubt upon the purpose of these data flows.

\parheading{Minecraft:} For all the age-related traces, we observed that all listed data type categories were collected by first parties (ATS and non-ATS) and shared with non-ATS third parties. For the child and adolescent traces, we observed all listed data type categories except personal identifiers being shared with third party ATS, which means device identifiers, personal characteristics, geolocation, user communications, and user interests and behaviors about users under 16 were observed being shared with third party ATS. The only difference between the adult and child/adolescent traces is that personal identifiers were not shared with third party ATS for the latter, otherwise they contain the same categories of data flows.

The privacy policy provided by Microsoft, which owns Minecraft, stated the following: ``we do not deliver personalized advertising to children whose birthdate in their Microsoft account identifies them as under 18 years of age''~\cite{microsoft_2023_ccpa_policy}. Since we observed data flows with identifiers and personal information shared with third party ATS in both the child and adolescent traces, we question the consistency of these behaviors and the privacy policy disclosure, since ATS are known to be related to personal advertising. Additionally, it is interesting that we observed sharing of all listed data type categories except personal identifiers with third party ATS for users under 16, but for the adult trace, personal identifiers was included. We are skeptical that simply removing the personal identifiers would result in de-identified data under CCPA, considering the other data types shared, and we question whether this was the intention.

\parheading{Quizlet:} We observed that all listed data type categories were collected by first parties, shared with third parties, and shared with third party ATS for all age traces. The lack of differentiation between the age traces and these collection and sharing behaviors overall are highly concerning for both the child and adolescent traces. We also observe the same behaviors for the logged-out trace, including all data type categories being shared with third party ATS, which raises concerns regarding their privacy policy disclosures.

The Quizlet privacy policy states the following for child users: ``We may use aggregated or de-identified information about children for research, analysis, marketing and other commercial purposes''~\cite{quizlet_privacy_policy_2021}. Thus, when consent is given for child users, they may share some personal information with third party ATS, which we have observed. However, we observed the same data type categories being shared with third party ATS while the user is logged out, which is problematic under COPPA and CCPA. While they may assert these data are de-identified and shared in aggregate, we observed all data type categories being shared and are skeptical that they are de-identified.

\parheading{Roblox:} The data flows for Roblox were largely the same across traces. The child, adolescent, and adult traces revealed all six listed data type categories being collected by first party non-ATS and shared with third party ATS. Additionally, all of the categories except geolocation were shared with third parties for all age-related traces. The only difference for the logged-out category compared to the age-related traces is that personal identifiers were not shared with non-ATS third parties and geolocation was not collected by non-ATS first parties. Thus, it seems this service is not significantly altering their behaviors by age or while the user is logged out. 

Roblox's privacy policy at the time these data were collected does not fully explain what we have observed. The child and adolescent data types shared with third party ATS are particularly concerning. Roblox stated the following in their privacy policy: ``We may share non-identifying data of all users regardless of their age'' for purposes such as marketing, reporting requirements, and service analytics~\cite{roblox_2023_privacy}. However, for the child and adolescent traces, we observed data types such as personal identifiers being shared with third party ATS, and we are skeptical that these data shared are ``non-identifying'' and are being used for the stated purposes considering the destinations include known third party ATS. Additionally, related to CCPA regulations, they stated that they ``have no actual knowledge of selling or sharing the Personal Information of minors under 16 years of age'' \cite{roblox_2023_privacy}. Considering what we have observed, we remain skeptical of these behaviors and policy disclosures.

\parheading{TikTok:} For the child and adolescent categories, we observed that all data type categories were collected by first parties, both non-ATS and ATS. Also, for both the child and adolescent traces, data flows including device identifiers and user communications were shared with both non-ATS and ATS third parties. The adolescent category also included data flows to third party ATS containing user interests and behaviors data. Comparing across age categories, the adult category contains more data flows to third parties, but overall the three age groups' observed data flows are very similar.

Considering the data types shared with third party ATS for the child and adolescent users, including device identifiers, user communications, and user interests and behaviors, TikTok's privacy policy stated the following: ``We may share the information that we collect with our corporate group or service providers as necessary for them to support the internal operations of the TikTok service''~\cite{tiktok_2023_children_priv}. However we are skeptical that ``service providers'' would include third party ATS, which are known to be related to non-functional advertising and tracking purposes.
Additionally, the policy stated, ``TikTok does not sell information from children to third parties and does not share such information with third parties for the purposes of cross-context behavioral advertising''~\cite{tiktok_2023_children_priv}. These statements lead us to question whether these observed practices of sharing data with third party ATS are for cross-context behavioral advertising or monetary consideration, since we observed both identifiers and personal information being shared with third party ATS.

\parheading{YouTube:} Both the regular YouTube service (\ie{} for adolescents and adults) and YouTube Kids are included in the YouTube category, and the child trace in Table \ref{table:platform_comparison_dataflows} corresponds to YouTube Kids specifically. We observed that all listed data type categories were collected by non-ATS first parties, and we observed no data flows to any third parties. Since Google owns YouTube and many ATS domains, it is understandable that we did not see any other third party domains contacted. The data flows in the table for the adolescent and adult groups were the same, with all listed data type categories collected by first party ATS. For the child trace, we observed that all listed data types except for personal identifiers and geolocation were collected by first party ATS.

In contrast with the other services discussed, YouTube's privacy policies seem to be consistent with the behaviors we observed. For example, the YouTube Kids privacy policy stated that they collect information including device type and settings, log information, and unique identifiers from users for various reasons, such as ``internal operational purposes'', ``personalized content'', and ``contextual advertising, including ad frequency capping'' \cite{youtubekids_2019_privacy}. It appears the observed behaviors are in line with the data collection and purposes disclosed for young users.

\parheading{Platform Differences:} Table \ref{table:platform_comparison_dataflows} also shows whether each data flow was observed on the mobile app, website, neither, or both platforms for each service. Since we were able to collect more clear text traffic on the websites compared to the mobile apps, it is understandable that we see far more website-only data flows than mobile-only, but we do see many data flows that were observed on both platforms. We observed data flows unique to the mobile platform only for Roblox, TikTok, Minecraft, and Duolingo, and these mobile-only data flows largely involved sharing data with third parties, including some third party ATS, such as personal and device identifiers, personal characteristics, geolocation, user communications, and user interests and behaviors. 
We observed many data flows unique to the website platform for all services and across all data types. These results suggest some potential development differences in mobile apps and their website counterparts that could lead to differences in what data is collected/shared about users on each platform.

\parheading{\textit{Key Takeaways:}} All but one of the services engaged in behaviors that raise concern for young users' privacy, including collecting and sharing both identifiers and personal information with third party ATS domains, which were often unclear and/or undisclosed in the services' privacy policies, while logged in and out. No service exhibited significantly different data processing treatment of the child and adolescent users compared to the adult users.

\subsection{Data Linkability}\label{sec:data_linkability_results}

We also investigate data linkability in the observed data flows per age group and while logged out. Data linkability refers to data types that are sent to the same third parties, including both ATS and non-ATS, that could be linked and used for tracking and profiling of users, as discussed in an SoK by Powar \etal{}~\cite{jovan_powar_sok_2023}. In our case, data linkability could occur if data flows containing at least one data type from both the identifiers and personal information categories are sent to the same third party (\eg{} a device hardware identifier and app/service usage data).

Across the entire dataset (not just linkable data flows), we found that the observed domains correspond to at least 212 companies (for some domains, we could not determine the owner). Based on our categorization of first/third parties and ATS domains, the destinations contacted include 
320 first parties (\eg{} \textit{tiktok.com}, \textit{roblox.com} and \textit{duolingo.com}), 
33 first party ATS (\eg{} \textit{browser.events.data.microsoft.com}, \textit{metrics.roblox.com}, and \textit{clarity.ms}), 
150 third parties (\eg{} \textit{cloudfront.net}, \textit{google\-apis.com}, and \textit{vimeocdn.com}), and 
485 third party ATS (\eg{} \textit{google-analytics.com}, \textit{doubleclick.net}, and \textit{amazon-adsystem.com}).

\begin{figure}[t!]
    \centering
    \includegraphics[width=\linewidth]{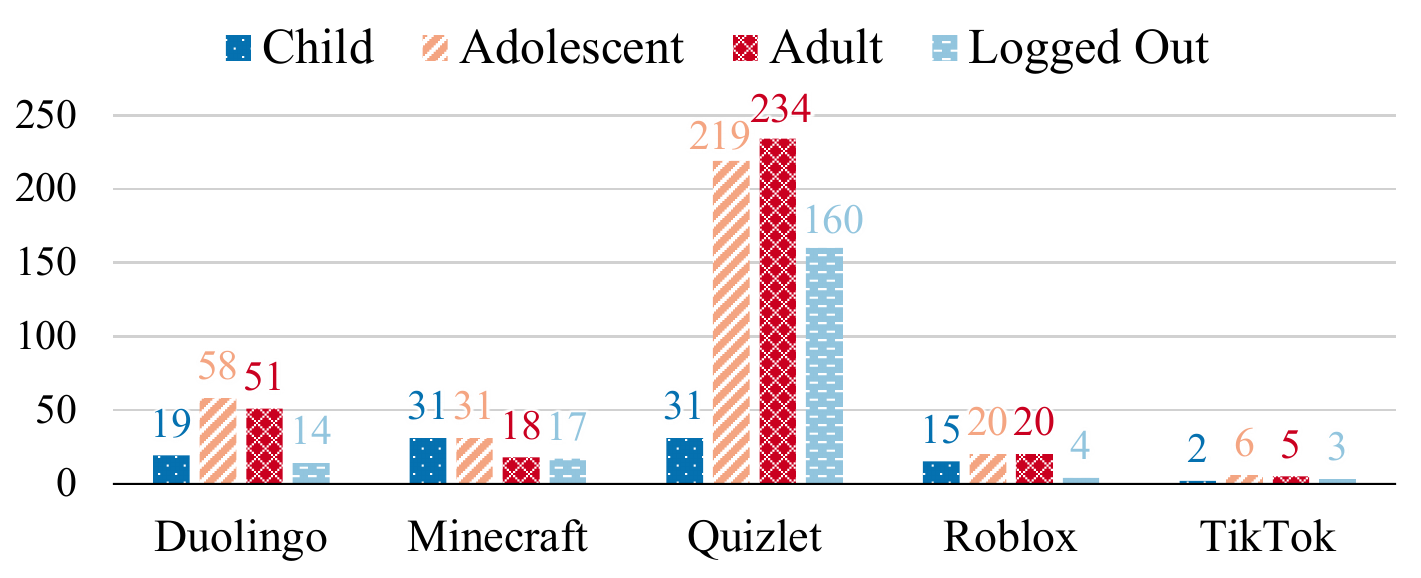}
    \caption{Counts of Third Parties Sent Linkable Data Types Per Service and Trace Category.
    {{\normalfont{Counts include third party domains, both ATS and non-ATS, that were sent linkable data types from each service per trace category (\ie{} child, adolescent, adult, and logged out). 
    }}}}
    \label{fig:data_linkage_domain_counts}
    \Description{Bar graph showing the counts of third parties that we observed were sent linkable data types per service and per trace category (child, adolescent, adult, and logged out).} 
\end{figure}

\begin{figure}[t!]
    \centering
    \includegraphics[width=\linewidth]{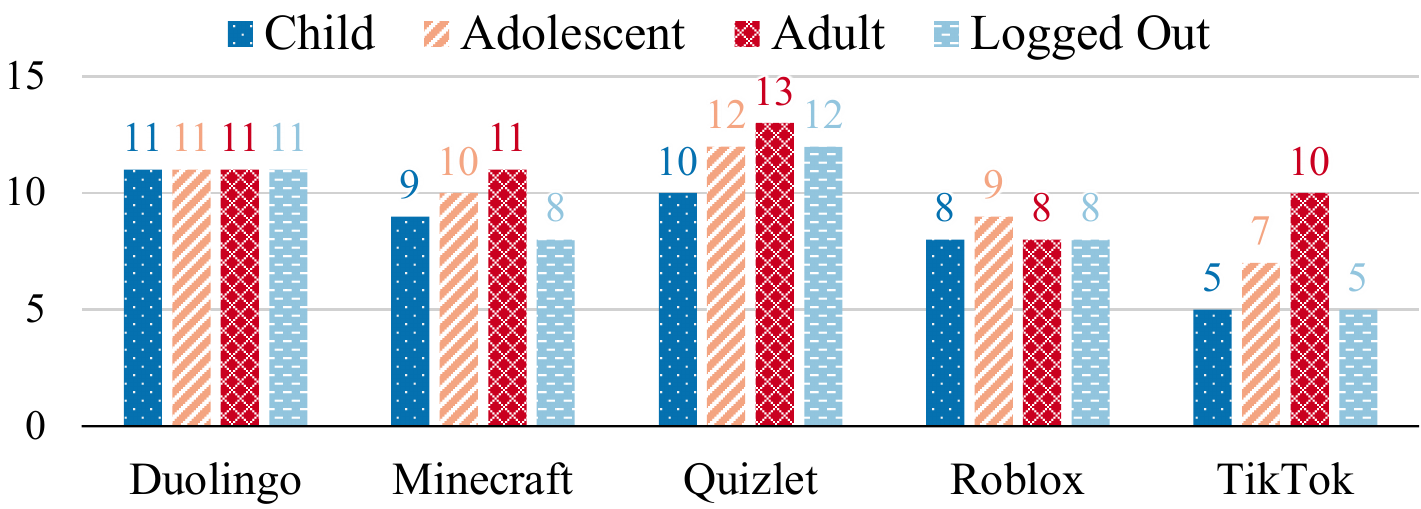}
    \caption{Sizes of Largest Sets of Linkable Data Types.
    {{\normalfont{A set of linkable data types is defined as all the data types that were shared with third party domains, including both ATS and non-ATS. The graph shows the size of the largest set of linkable data types shared by each service per trace category (\ie{} child, adolescent, adult, and logged out). 
    }}}}
    \label{fig:data_linkage_set_sizes}
    \Description{Bar graph showing the largest sets of linkable data types sent to third parties per service and per trace category (child, adolescent, adult, and logged out).} 
\end{figure}

Figure~\ref{fig:data_linkage_domain_counts} shows the counts of third parties that were specifically sent linkable data types based on the observed network traffic per service and per trace category. We observe that Quizlet had the highest counts of third parties to whom they sent linkable data for all categories except for the child trace, and YouTube has the least (all counts are zero), as YouTube only contacted first parties in our traces. We also observed a trend in the counts of third parties when comparing the age groups, with most of the services sharing linkable data types with a smaller number of third parties for the child category compared to that of the adolescent and adult categories. However, we observed high counts for the adolescent category similar to those of the adult across services (\eg{} 219 for adolescent vs. 234 for adult in Quizlet).

\begin{figure}[t!]
    \centering
    \includegraphics[width=\linewidth]{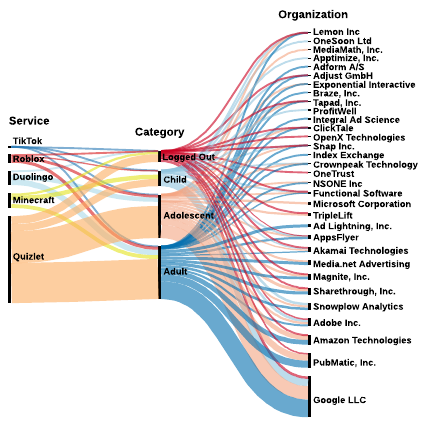}
    \caption{Most Frequent Third Party ATS Domains Sent Linkable Data Types.
    {{\normalfont{Alluvial diagram visualizes the top-10 most contacted third party ATS, shown based on their organizations, that were sent linkable data types by each service per trace category (\ie{} child, adolescent, adult, and logged out). 
    }}}}
    \label{fig:ats_linkage_alluvial}
    \Description{Alluvial diagram showing the flow of data from each service per trace category (child, adolescent, adult, and logged out) to the most frequently contacted advertising and tracking organizations.} 
\end{figure}

We also analyze the sets of linkable data types transmitted by the services, and we show the sizes of the largest sets per trace and service in Figure~\ref{fig:data_linkage_set_sizes}. The largest set of linkable data types out of the entire dataset was shared by Quizlet in the adult trace, which included 13 (out of 19 total observed in the entire dataset) data types: network connection information, language, device information, app or service usage, service information, products and advertising, account settings, aliases, name, login information, location time, device software identifiers, and reasonably linkable personal identifiers. The most common set of linkable data types shared across the entire dataset included 5 data types: network connection information, language, service information, app or service usage, and service information.

To give context for the particular ATS with whom these services shared linkable data per trace category, Figure~\ref{fig:ats_linkage_alluvial} shows an alluvial diagram of the most contacted third party ATS domains per service, mapped with the organizations that own those domains, with whom linkable data was shared. Due to space, we only show the top 10 most contacted ATS per service and per category, since there are over 190 third party ATS domains. The bar widths indicate the frequency of contact with those domains, demonstrating the relative difference in frequency of linkable data flows between services, with Quizlet contacting the most ATS domains with linkable data flows most frequently, and TikTok contacting the least for both. The 32 organizations shown include Google, Pubmatic, Amazon, Adobe, and Microsoft.

\parheading{\textit{Key Takeaways:}} All services except one sent linkable data types to third party domains, including ATS and non-ATS, for all age groups and while logged out. 
These services sent multiple linkable data types to many of the same third party domains, demonstrating a lack of differentiation between age groups and while logged out, as well as risky linkable data sharing behaviors overall.

%% file: tables/web_vs_mobile_data_flows.tex
\newcommand\rot{\rotatebox{45}}
\newcommand\YES{$\bullet$}
\newcommand\NO{\textemdash}

\newcommand\mobile{\faIcon{mobile-alt}}
\newcommand\website{\faIcon{mouse-pointer}}
\newcommand\both{$\bullet$}

\begin{table*}[t]
  \centering
  \small 
  \caption{Data Flows Observed by Age Category for Website and Mobile Platforms.
  {\normalfont{Data flows are indicated by the presence of a symbol in the grid between a data type category and destination. The age-specific flows are merged from the account creation and logged-in traces. ``Collect'' indicates data sent to first parties. ``Share'' indicates data sent to third parties.}}}
  \begin{tabular}{llp{0.01\linewidth}p{0.01\linewidth}p{0.01\linewidth}p{0.01\linewidth}p{0.01\linewidth}p{0.01\linewidth}p{0.01\linewidth}p{0.01\linewidth}p{0.01\linewidth}p{0.01\linewidth}p{0.01\linewidth}p{0.01\linewidth}p{0.01\linewidth}p{0.01\linewidth}p{0.01\linewidth}p{0.01\linewidth}}
    & &   \multicolumn{4}{c}{\bf{Child}} & \multicolumn{4}{c}{\bf{Adolescent}} & \multicolumn{4}{c}{\bf{Adult}} & \multicolumn{4}{c}{\bf{Logged Out}} \\\hhline{~~----------------}
    \bf{Service} & \multicolumn{1}{c}{\bf{Data Type}} &    \rot{Collect 1st} & \rot{Collect 1st ATS} & \rot{Share 3rd} & \rot{Share 3rd ATS} &    
                                                \rot{Collect 1st} & \rot{Collect 1st ATS} & \rot{Share 3rd} & \rot{Share 3rd ATS} & 
                                                \rot{Collect 1st} & \rot{Collect 1st ATS} & \rot{Share 3rd} & \rot{Share 3rd ATS} & 
                                                 \rot{Collect 1st} & \rot{Collect 1st ATS} & \rot{Share 3rd} & \rot{Share 3rd ATS} \\\hline

\multicolumn{1}{c|}{}  &   Personal Identifiers & \multicolumn{1}{|c}{\both} & \NO & \website  & \multicolumn{1}{c|}{\both}  & \both & \NO & \website  & \multicolumn{1}{c|}{\both}  & \both & \NO  & \website  &  \multicolumn{1}{c|}{\both}  & \both & \NO & \NO & \mobile \\
\multicolumn{1}{c|}{}  &   Device Identifiers & \multicolumn{1}{|c}{\both} & \NO & \both  & \multicolumn{1}{c|}{\both}  & \both & \NO & \both  & \multicolumn{1}{c|}{\both}  & \both & \NO  & \both  &  \multicolumn{1}{c|}{\both}  & \both & \NO & \both & \both \\
\multicolumn{1}{c|}{Duolingo}  &   Personal Characteristics & \multicolumn{1}{|c}{\both} & \NO & \website  & \multicolumn{1}{c|}{\both}  & \both & \NO & \website  & \multicolumn{1}{c|}{\both}  & \both & \NO  & \website  &  \multicolumn{1}{c|}{\both}  & \both & \NO & \website & \both \\
\multicolumn{1}{c|}{}  &   Geolocation & \multicolumn{1}{|c}{\both} & \NO & \NO  & \multicolumn{1}{c|}{\both}  & \both & \NO & \NO  & \multicolumn{1}{c|}{\both}  & \both & \NO  & \NO  &  \multicolumn{1}{c|}{\both}  & \both & \NO & \NO & \mobile \\
\multicolumn{1}{c|}{}  &   User Communications & \multicolumn{1}{|c}{\both} & \NO & \both  & \multicolumn{1}{c|}{\both}  & \both & \NO & \both  & \multicolumn{1}{c|}{\both}  & \both & \NO  & \both  &  \multicolumn{1}{c|}{\both}  & \both & \NO & \both & \both \\
\multicolumn{1}{c|}{}  &   User Interests and Behaviors & \multicolumn{1}{|c}{\both} & \NO & \both  & \multicolumn{1}{c|}{\both}  & \both & \NO & \both  & \multicolumn{1}{c|}{\both}  & \both & \NO  & \both  &  \multicolumn{1}{c|}{\both}  & \both & \NO & \both & \both \\\hline

\multicolumn{1}{c|}{}  &   Personal Identifiers & \multicolumn{1}{|c}{\both} & \both & \mobile  & \multicolumn{1}{c|}{\NO}  & \both & \both & \mobile  & \multicolumn{1}{c|}{\NO}  & \both & \both  & \mobile  &  \multicolumn{1}{c|}{\mobile}  & \mobile & \website & \NO & \NO \\
\multicolumn{1}{c|}{}  &   Device Identifiers & \multicolumn{1}{|c}{\both} & \both & \both  & \multicolumn{1}{c|}{\both}  & \both & \both & \both  & \multicolumn{1}{c|}{\both}  & \both & \both  & \both  &  \multicolumn{1}{c|}{\both}  & \both & \both & \website & \both \\
\multicolumn{1}{c|}{Minecraft}  &   Personal Characteristics & \multicolumn{1}{|c}{\both} & \both & \both  & \multicolumn{1}{c|}{\both}  & \both & \both & \both  & \multicolumn{1}{c|}{\both}  & \both & \both  & \both  &  \multicolumn{1}{c|}{\both}  & \both & \website & \website & \both \\
\multicolumn{1}{c|}{}  &   Geolocation & \multicolumn{1}{|c}{\both} & \website & \website  & \multicolumn{1}{c|}{\mobile}  & \website & \website & \website  & \multicolumn{1}{c|}{\mobile}  & \both & \website  & \website  &  \multicolumn{1}{c|}{\mobile}  & \mobile & \website & \NO & \mobile \\
\multicolumn{1}{c|}{}  &   User Communications & \multicolumn{1}{|c}{\both} & \both & \both  & \multicolumn{1}{c|}{\both}  & \both & \both & \both  & \multicolumn{1}{c|}{\both}  & \both & \both  & \both  &  \multicolumn{1}{c|}{\both}  & \both & \both & \website & \both \\
\multicolumn{1}{c|}{}  &   User Interests and Behaviors & \multicolumn{1}{|c}{\both} & \both & \website  & \multicolumn{1}{c|}{\both}  & \both & \both & \both  & \multicolumn{1}{c|}{\both}  & \both & \both  & \website  &  \multicolumn{1}{c|}{\both}  & \both & \both & \website & \both \\\hline

\multicolumn{1}{c|}{}  &   Personal Identifiers & \multicolumn{1}{|c}{\both} & \NO & \both  & \multicolumn{1}{c|}{\website}  & \both & \NO & \both  & \multicolumn{1}{c|}{\both}  & \both & \NO  & \both  &  \multicolumn{1}{c|}{\both}  & \website & \NO & \both & \both \\
\multicolumn{1}{c|}{}  &   Device Identifiers & \multicolumn{1}{|c}{\both} & \NO & \both  & \multicolumn{1}{c|}{\both}  & \both & \NO & \both  & \multicolumn{1}{c|}{\both}  & \both & \NO  & \both  &  \multicolumn{1}{c|}{\both}  & \both & \NO & \both & \both \\
\multicolumn{1}{c|}{Quizlet}  &   Personal Characteristics & \multicolumn{1}{|c}{\both} & \NO & \both  & \multicolumn{1}{c|}{\both}  & \both & \NO & \both  & \multicolumn{1}{c|}{\both}  & \both & \NO  & \both  &  \multicolumn{1}{c|}{\both}  & \both & \NO & \both & \both \\
\multicolumn{1}{c|}{}  &   Geolocation & \multicolumn{1}{|c}{\website} & \NO & \both  & \multicolumn{1}{c|}{\both}  & \website & \NO & \both  & \multicolumn{1}{c|}{\both}  & \website & \NO  & \both  &  \multicolumn{1}{c|}{\both}  & \website & \NO & \both & \both \\
\multicolumn{1}{c|}{}  &   User Communications & \multicolumn{1}{|c}{\both} & \NO & \both  & \multicolumn{1}{c|}{\both}  & \both & \NO & \both  & \multicolumn{1}{c|}{\both}  & \both & \NO  & \both  &  \multicolumn{1}{c|}{\both}  & \both & \NO & \both & \both \\
\multicolumn{1}{c|}{}  &   User Interests and Behaviors & \multicolumn{1}{|c}{\both} & \NO & \both  & \multicolumn{1}{c|}{\both}  & \both & \NO & \both  & \multicolumn{1}{c|}{\both}  & \both & \NO  & \both  &  \multicolumn{1}{c|}{\both}  & \both & \NO & \both & \both \\\hline

\multicolumn{1}{c|}{}  &   Personal Identifiers & \multicolumn{1}{|c}{\both} & \both & \mobile  & \multicolumn{1}{c|}{\website}  & \both & \both & \mobile  & \multicolumn{1}{c|}{\website}  & \both & \both  & \mobile  &  \multicolumn{1}{c|}{\website}  & \website & \website & \NO & \website \\
\multicolumn{1}{c|}{}  &   Device Identifiers & \multicolumn{1}{|c}{\both} & \both & \both  & \multicolumn{1}{c|}{\both}  & \both & \both & \both  & \multicolumn{1}{c|}{\both}  & \both & \both  & \both  &  \multicolumn{1}{c|}{\both}  & \both & \both & \website & \website \\
\multicolumn{1}{c|}{Roblox}  &   Personal Characteristics & \multicolumn{1}{|c}{\both} & \both & \both  & \multicolumn{1}{c|}{\both}  & \both & \both & \both  & \multicolumn{1}{c|}{\both}  & \both & \both  & \both  &  \multicolumn{1}{c|}{\both}  & \both & \both & \website & \website \\
\multicolumn{1}{c|}{}  &   Geolocation & \multicolumn{1}{|c}{\website} & \NO & \NO  & \multicolumn{1}{c|}{\website}  & \website & \NO & \NO  & \multicolumn{1}{c|}{\both}  & \website & \NO  & \NO  &  \multicolumn{1}{c|}{\website}  & \NO & \NO & \NO & \website \\
\multicolumn{1}{c|}{}  &   User Communications & \multicolumn{1}{|c}{\both} & \both & \both  & \multicolumn{1}{c|}{\both}  & \both & \both & \both  & \multicolumn{1}{c|}{\both}  & \both & \both  & \both  &  \multicolumn{1}{c|}{\both}  & \both & \both & \website & \website \\
\multicolumn{1}{c|}{}  &   User Interests and Behaviors & \multicolumn{1}{|c}{\both} & \both & \both  & \multicolumn{1}{c|}{\both}  & \both & \both & \both  & \multicolumn{1}{c|}{\both}  & \both & \both  & \both  &  \multicolumn{1}{c|}{\both}  & \both & \website & \website & \website \\\hline

\multicolumn{1}{c|}{}  &   Personal Identifiers & \multicolumn{1}{|c}{\website} & \website & \NO  & \multicolumn{1}{c|}{\NO}  & \website & \website & \website  & \multicolumn{1}{c|}{\NO}  & \website & \website  & \website  &  \multicolumn{1}{c|}{\mobile}  & \website & \website & \NO & \NO \\
\multicolumn{1}{c|}{}  &   Device Identifiers & \multicolumn{1}{|c}{\both} & \both & \website  & \multicolumn{1}{c|}{\mobile}  & \both & \both & \website  & \multicolumn{1}{c|}{\mobile}  & \both & \both  & \website  &  \multicolumn{1}{c|}{\mobile}  & \both & \website & \website & \mobile \\
\multicolumn{1}{c|}{TikTok}  &   Personal Characteristics & \multicolumn{1}{|c}{\website} & \website & \website  & \multicolumn{1}{c|}{\NO}  & \website & \website & \website  & \multicolumn{1}{c|}{\NO}  & \website & \website  & \website  &  \multicolumn{1}{c|}{\mobile}  & \website & \website & \website & \NO \\
\multicolumn{1}{c|}{}  &   Geolocation & \multicolumn{1}{|c}{\website} & \website & \NO  & \multicolumn{1}{c|}{\NO}  & \website & \website & \NO  & \multicolumn{1}{c|}{\NO}  & \website & \website  & \NO  &  \multicolumn{1}{c|}{\mobile}  & \website & \website & \NO & \NO \\
\multicolumn{1}{c|}{}  &   User Communications & \multicolumn{1}{|c}{\both} & \both & \website  & \multicolumn{1}{c|}{\mobile}  & \both & \both & \website  & \multicolumn{1}{c|}{\mobile}  & \both & \both  & \website  &  \multicolumn{1}{c|}{\mobile}  & \both & \website & \website & \mobile \\
\multicolumn{1}{c|}{}  &   User Interests and Behaviors & \multicolumn{1}{|c}{\website} & \website & \website  & \multicolumn{1}{c|}{\NO}  & \website & \website & \website  & \multicolumn{1}{c|}{\mobile}  & \website & \website  & \website  &  \multicolumn{1}{c|}{\mobile}  & \both & \website & \website & \NO \\\hline

\multicolumn{1}{c|}{}  &   Personal Identifiers & \multicolumn{1}{|c}{\website} & \NO & \NO  & \multicolumn{1}{c|}{\NO}  & \both & \website & \NO  & \multicolumn{1}{c|}{\NO}  & \website & \website  & \NO  &  \multicolumn{1}{c|}{\NO}  & \website & \NO & \NO & \NO \\
\multicolumn{1}{c|}{}  &   Device Identifiers & \multicolumn{1}{|c}{\website} & \website & \NO  & \multicolumn{1}{c|}{\NO}  & \both & \website & \NO  & \multicolumn{1}{c|}{\NO}  & \both & \website  & \NO  &  \multicolumn{1}{c|}{\NO}  & \website & \website & \NO & \NO \\
\multicolumn{1}{c|}{YouTube}  &   Personal Characteristics & \multicolumn{1}{|c}{\website} & \website & \NO  & \multicolumn{1}{c|}{\NO}  & \website & \website & \NO  & \multicolumn{1}{c|}{\NO}  & \website & \website  & \NO  &  \multicolumn{1}{c|}{\NO}  & \website & \website & \NO & \NO \\
\multicolumn{1}{c|}{}  &   Geolocation & \multicolumn{1}{|c}{\website} & \NO & \NO  & \multicolumn{1}{c|}{\NO}  & \both & \website & \NO  & \multicolumn{1}{c|}{\NO}  & \website & \website  & \NO  &  \multicolumn{1}{c|}{\NO}  & \website & \website & \NO & \NO \\
\multicolumn{1}{c|}{}  &   User Communications & \multicolumn{1}{|c}{\website} & \website & \NO  & \multicolumn{1}{c|}{\NO}  & \both & \website & \NO  & \multicolumn{1}{c|}{\NO}  & \both & \website  & \NO  &  \multicolumn{1}{c|}{\NO}  & \website & \website & \NO & \NO \\
\multicolumn{1}{c|}{}  &   User Interests and Behaviors & \multicolumn{1}{|c}{\website} & \website & \NO  & \multicolumn{1}{c|}{\NO}  & \both & \website & \NO  & \multicolumn{1}{c|}{\NO}  & \both & \website  & \NO  &  \multicolumn{1}{c|}{\NO}  & \website & \website & \NO & \NO \\\hline

  \end{tabular}
    \begin{tabular}{ll}\small
    {\both} : data flow observed on both website and mobile &
    {\NO} : data flow not observed on either platform \\
    {\website} : data flow observed only on website platform &
    {\mobile} : data flow observed only on mobile platform \\
    \\
    \end{tabular}

  \label{table:platform_comparison_dataflows}
\end{table*}

%% file: 5_discussion.tex
\section{Discussion}\label{discussion}

\subsection{Privacy Implications} All the services we investigated in this work with {\diffaudit} collected and/or shared personal information about child and adolescent users in such a way that raises concerns with respect to COPPA and CCPA.
When comparing between the child, adolescent, and adult trace data flows with our differential analysis approach, we found that these services do not appear to be altering their data collection and sharing practices as much as is expected for child and adolescent users. 
Every service also collected and/or shared information about users prior to knowing their age and obtaining consent (\ie{} while logged out), including with third party ATS, which is inappropriate under COPPA and CCPA regulations. General audience services should wait until they are informed of user age and receive the appropriate consent so that they proceed with the appropriate data collection and sharing for the age category.
Also, all but one of the services shared data with third parties that could enable data linkage of children and adolescent users.
Finally, all but one of the services had privacy policies that were inconsistent with the data flows we observed for the child and adolescent traces.

Our findings complement the work of Reyes \etal{}~\cite{irwin_reyes_wont_2018}, which is the closest work to our study. Reyes \etal{} investigated COPPA compliance of Android apps for children and found that a majority of children's apps studied had potential COPPA compliance issues largely because of third-party SDKs used in the apps. Also, 19\% of the apps they studied collected identifiers and personal information through SDKs. However, as we discussed in Section~\ref{sec:methods_data_types}, our work identifies considerably more data types (up to 35 based on our ontology), whereas \cite{irwin_reyes_wont_2018} focused on a smaller set of data types (\eg{} device identifiers, contact information, and geolocation). Additionally, we are able to explore a richer set of data collection and sharing practices because we study both mobile apps and websites in context of COPPA and CCPA, further advancing our knowledge about how online services handle children's and adolescents data.

\subsection{Recommendations} Based on our observations, children's and adolescents' online privacy are not being adequately protected. General audience services need to improve their practices to protect their young users  and ensure they are compliant with COPPA and CCPA. We urge service providers to scrutinize every piece of data being collected and shared before and after users log in to their service. Service providers should also be more transparent in their privacy policies and correctly disclose their data collection and sharing behaviors to ensure users and parents are able to provide informed consent.

\subsection{Limitations \& Future Directions}

{\em Automation.} There are limitations in the usage of machine learning to automatically classify data types from network traffic data and in analyzing network traffic, which is not entirely clear-text.  We will continue to evolve our data type classification as large language models continue to improve.
In this study, we also chose to manually interact with the services to increase the quality and depth of our dataset. However, even if the interaction with the services becomes automated, the {\diffaudit} framework can easily be applied.

{\em Other Applications.} At the time of this study, there were only six general audience services among the top-100 most popular that fit our criteria: Duolingo, Minecraft, Quizlet, Roblox, TikTok, and YouTube/YouTube Kids. However, our {\diffaudit} framework can be applied to other general audience services, as they become available in the future, since {\diffaudit} relies only on the availability of network traffic. Furthermore, since {\diffaudit} is platform-agnostic and relies only on network traffic, it can be applied to audit data collection and sharing practices of online services for young users on emerging platforms, such as virtual reality applications, voice assistants, and smart toys. Our differential analysis approach can also be re-purposed for other age-based auditing, such as content analysis and dark patterns.

{\em Using {\diffaudit}.} We envision that {\diffaudit} can be used by researchers and regulators to identify potentially problematic behaviors and to audit compliance of general audience online services with COPPA and CCPA.
We plan to make {\diffaudit}'s implementation and datasets available.

%% file: 6_conclusion.tex
\section{Conclusion}\label{conclusion}

In this work, we present {\diffaudit}, a platform-agnostic auditing framework for general audience services to investigate privacy behaviors and regulatory compliance. 
Within this framework, we construct data flows from network traffic through destination analysis and our novel data type classification method with GPT-4 and our data type ontology, which identifies considerably more data types than prior work. 
We conduct differential analyses to investigate the observed data flows generated by child, adolescent, and adult user profiles. We compare the data flows against regulations and privacy policies, analyze the data flows while logged in and out (\ie{} before and after consent is given and user age is disclosed), and investigate data linkage.
We apply {\diffaudit} to popular online services on mobile and website platforms and find that all audited services engage in problematic behaviors for children and adolescents, including collecting and/or sharing data while users are logged out, inconsistent privacy policy disclosures, and data flows shared with third parties that enable data linkage for children and adolescents.
{\diffaudit} enables auditors to investigate network traffic through  differential analysis, utilizing an automated and customizable data type classification method, to identify suspicious behaviors in network traffic for any platform with a focus on children's and adolescents' privacy under COPPA and CCPA.

%% file: 7_acknowledgements.tex
\section{Acknowledgements}
This work has been supported by NSF awards FG22924 and FG22490.

%% file: appendix.tex
\appendix

\section*{Appendices}

\input{tables/full_data_category_ontology}

\section{Ethics Considerations}\label{sec:ethics} 
We created test accounts on online services and collected network traffic while the authors interacted with those services. This may lead to advertisers losing some ad revenue, since ads may show up as we use the services in our data collection sessions, but we only use each service as long as is required to exhaust all the functionalities, which is not long, thus minimizing our impact. However, no real users' data were collected or used. We plan to reach out to the online service providers and notify them of our observations and findings (responsible disclosure).

\section{Data Type Ontology}\label{app:data_type_categories}
In Section~\ref{post_process}, we discuss how we label data types according to our data type ontology categories. In this appendix, we present the complete data type ontology in Table~\ref{table:datatype_categories}.

\section{Data Type Classification Methods}\label{app:labeling_methods}
In Section~\ref{post_process}, we discuss data type classification methods, including using OpenAI's GPT-4 and experiments with alternative methods. In this appendix, we further discuss these methods.

\subsection{OpenAI's GPT-4}
For our GPT-4 approach, we used OpenAI's GPT-4 8K context model \cite{gpt-4_2023} with their Chat Completion API \cite{chat_completions_api_2023}, which can take in prompts and generate textual responses. 
A challenge with using GPT-4 is designing the input prompt, since the prompt is what guides the model in understanding and completing the given task and can drastically change the output depending on how the task is worded and what information is given to the model. 
We experimented with several iterations of prompts and followed OpenAI's GPT best practices guide \cite{gpt_2023_best_practices}. 
We attempted to design a prompt that resulted in a reliable output format, which we required to be able to operationalize the output of the model for classification, and that yielded high accuracy. 
We found that asking the model to provide a confidence score from 0 to 1 for each classification and to explain its reasoning yielded better performance. 
Below is the final prompt used for data type classification using GPT-4 Chat Completions:
\begin{adjustwidth}{2em}{2em}
``You are a text classifier for network traffic payload data. I am going to give you some categories and examples for each category. Then I will give you text sequences that I want you to categorize using the provided categories. The input texts were collected from network traffic payloads. Try to determine the meaning of the input texts and use the similarity of the categories and input texts to do the classification. For text with acronyms and abbreviations, use the meaning of the acronyms and abbreviations to do the classification. Provide an explanation for each classification in 15 words or less. Report a score of confidence on a scale of 0 to 1 for each categorization. Format your response exactly like this for each input text: <input text> // <category> // <score> // <explanation>.''
\end{adjustwidth}

\subsection{Classifier Experiments}
For the fuzzy string matching method, we utilized the examples in each category as the strings to which the model would attempt to match a given input string. If a given input string matched to an example, the category for that example was used as the label for that data type. We utilized the library PolyFuzz \cite{maarten_grootendorst_maartengrpolyfuzz_2021}, which enables customizable fuzzy string matching with Hugging Face transformers as well as word embedding methods such as TF-IDF. We experimented with the TF-IDF and BERT models. For the zero-shot classification approach, we used the zero-shot classification pipeline \cite{huggingface_zeroshot_2023} provided by Hugging Face with bart-large-mnli \cite{facebookbart-large-mnli_2023}, which is based on the BART model \cite{mike_lewis_bart_2019}. We only inputted the data type categories, and not any of the examples, as labels for classification. Finally, for the few-shot classification approach, we utilized SetFit \cite{tunstall_efficient_2022}, which is a few-shot learning framework that does not require any prompts, avoiding high variability that other few-shot learning methods suffer from due to reliance on prompts. We inputted our categories and examples as the labeled training data and utilized the one-vs-rest strategy as we are conducting multi-label classification.

%% file: tables/full_data_category_ontology.tex
\newcommand\lwid{1.75cm}
\newcommand\mwid{1.5cm}
\newcommand\nwid{4cm}

\begin{table*}[t!]
  \centering
  \footnotesize
  \caption{Data Type Ontology for Data Type Classification based on COPPA and CCPA.
  \normalfont{{Level 3 is used for data type classification labels. Some labels have been grouped into level 4 due to space. See Table~\ref{table:data_type_35_categories} for all 35 labels.}}}
  \begin{tabular}{|p{\mwid}|p{\lwid}|p{\nwid}|p{0.45\linewidth}|}\hline
    \bf{Level 1} & \bf{Level 2} & \bf{Level 3} & \bf{Level 4} \\\hline
    \multirow{8}{\mwid}{Identifiers \cite{ccpa_defn_california_2018, coppa_defn_childrens_2013}} &  \multirow{6}{\lwid}{Personal Identifiers} &  Name & first and last name, user name \\\hhline{~~--}
                                 &                                        &  Linked Personal Identifiers  & social security number, driver's license number, state identification card number, passport number\\\hhline{~~--}
                                 &                                        &  Contact Information &  email address, telephone number, phone number \\\hhline{~~--}
                                 &                                        &  Reasonably Linkable Personal Identifiers & IP address, unique pseudonym/alias \\\hhline{~~--}
                                 &                                        &  Aliases & alias, online identifier, unique personal identifier, unique id, GUID (globally unique identifier), UUID (universally unique identifier) \\\hhline{~~--}
                                 &                                        &  Customer Numbers  & customer number, account name, insurance policy number, bank account number, credit card number, debit card number \\\hhline{~~--}
                                 &                                        & Login Information & password, login, authorization, authentication \\\hhline{~---}
                                 & \multirow{3}{\lwid}{Device Identifiers}&  Device Hardware Identifiers & IMEI (international mobile equipment identity), MAC (media access control) address, unique device identifier, processor serial number, device serial number \\\hhline{~~--}
                                 &                                        &  Device Software Identifiers & advertising identifier, cookie, pixel tag, beacon, tracking identifier \\\hhline{~~--}
                                 &                                        & Device Information &  display, height, width, FPS (frames per second), browser, bitrate, ABR (adaptive bitrate), ABR bitrate map, speed, device, delay, OS (operating system), rate, screen, sound, memory, history, CPU (central processing unit), buffer, latency, download, load, frame, depth, download speed, render \\\hline

    \multirow{16}{\mwid}{Personal Information \cite{ccpa_defn_california_2018, coppa_defn_childrens_2013}}  &   \multirow{2}{\lwid}{Personal Characteristics}   &   Protected Classifications  & race, skin color, national origin, ancestry, age, language, religion, gender, sexual orientation, marital status, military or veteran status, medical conditions, genetic information, disabilities \\\hhline{~~--}
                                            &                                               &   Biometric Information &  DNA, images, voiceprint, patterns, rhythms, physical characteristics or descriptions \\\hhline{~---}
                                            &  Personal History       &   Personal History  & employment, education, financial information, medical information \\\hhline{~---}
                                            &  \multirow{2}{\lwid}{Geolocation}             &   Precise Geolocation & GPS (global positioning system) location, coordinates, postal address, latitude, longitude \\\hhline{~~--}
                                            &                                               &   Coarse Geolocation  &   city, town, country, region \\\hhline{~~--}
                                            &                                               &   Location Time   & time, timestamp, timezone, time offset, date \\\hhline{~---}
                                            &  \multirow{4}{\lwid}{User Communications}     &   Communications  & audio, text, and video communications \\\hhline{~~--}
                                            &                                               &   Contacts  & people communication with, contact list \\\hhline{~~--}
                                            &                                               &   Internet Activity   & IP addresses communicated with, browsing history, search history \\\hhline{~~--}
                                            &                                               & Network Connection Information &  request, response, DNS (domain name system), TCP (transmission control protocol), TLS (transport layer security), RTT (round trip time), TTFB (time to first byte), protocol, client, connection, key, payload, host, referer, telemetry, cache \\\hhline{~---}
                                            &   \multirow{1}{\lwid}{Sensors}                &   Sensor Data & audio recordings, video recordings, other sensor data (e.g., thermal, olfactory) \\\hhline{~---}
                                            &  \multirow{5}{\lwid}{User Interests and Behavior} &   Products and Advertising & records of personal property, products or services, or those considered, consumer's interaction with an advertisement, advertisement/ad engagement, bid, analytics, marketing, third party, advertiser \\\hhline{~~--}
                                            &                                               &   App or Service Usage    & user interaction with an application, user interaction with a website, session, usage session, content, video, audio, video buffer, audio buffer, play, volume, avatar, behavior, action, event, data, status, duration, timing \\\hhline{~~--}
                                            &                                               & Account Settings & account, settings, consent, permission\\\hhline{~~--}
                                            &                                               & Service Information & server, SDK (software development kit), API (application programming interface), site, URL (uniform resource locator), domain, version, script, URI (uniform resource identifier), application, page, app, CDN (content delivery network), DOM (document object model) \\\hhline{~~--}
                                            &                                               &  Inferences &   user preferences, characteristics, psychological trends, predispositions, behavior, attitudes, intelligence, abilities, aptitudes, personality, purchase history, purchase tendency \\\hline

  \end{tabular}
    
  \label{table:datatype_categories}
\end{table*}